\documentclass[10pt,twocolumn]{article}  
\usepackage{ol2}
\usepackage{graphicx,bm,psfrag} 
\usepackage{epstopdf}  
\newcommand{\ket}[1]{|#1\rangle}             
\begin{document}

\twocolumn[  
\title{Time-division multiplexing of the orbital angular momentum of light}
\author{Ebrahim Karimi$^1$, Lorenzo Marrucci$^{1,3}$, Corrado de Lisio$^{1,3}$ and Enrico Santamato$^{1,2}$}
\affiliation{$^1$Dipartimento di Scienze Fisiche, Universit\`{a} di Napoli ``Federico II'', Complesso di Monte S. Angelo, 80126 Napoli, Italy}
\affiliation{$^2$CNISM-Consorzio Nazionale Interuniversitario per le Scienze
Fisiche della Materia, Napoli, Italy}
\affiliation{$^3$CNR-SPIN, Complesso Universitario di Monte S. Angelo, 80126 Napoli, Italy}
\begin{abstract}
We present an optical setup for generating a sequence of light
pulses in which the orbital angular momentum (OAM) degree of freedom
is correlated with the temporal one. The setup is based on a single
$q$-plate within a ring optical resonator. By this approach, we demonstrate the
generation of a train of pulses carrying increasing values of OAM,
or, alternatively, of a controlled temporal sequence of pulses having
prescribed OAM superposition states. Finally, we exhibit an
``OAM-to-time conversion'' apparatus dividing different input OAM
states into different time-bins. The latter application provides a simple
approach to digital spiral spectroscopy of pulsed light.
\end{abstract}
] 

The orbital angular momentum (OAM) of light has been attracting an increasing
interest in the last years, owing both to fundamental reasons and to its potential
for multi-valued encoding of information, both in a classical and quantum regime
\cite{frankearnold08,yao11}. One key step in fully unreleasing this potential
is to develop a convenient photonic technology for OAM generation, manipulation, and
detection. The current OAM technology, e.g. including holograms and spatial light
modulators (SLM) \cite{mair01}, spiral phase-plates \cite{oemrawsing04}, Dove prisms
in suitable interferometers \cite{leach02}, etc., has important limitations in terms of
ease of use, switching speed, cost, etc., so that the need for new solutions
remains strong (see, e.g., \cite{berkhout10}). In this paper, we propose a novel
approach to OAM generation and detection based on setting up
a correlation between this degree of freedom and the temporal one,
for example within a regular sequence of optical pulses. This correlation
can in turn be exploited to generate a time-to-OAM  transfer of the
information encoded in the optical field, or vice versa, depending on
the details of the scheme.

\begin{figure}[tb]
\begin{center}
    \includegraphics[width=8cm,draft=false]{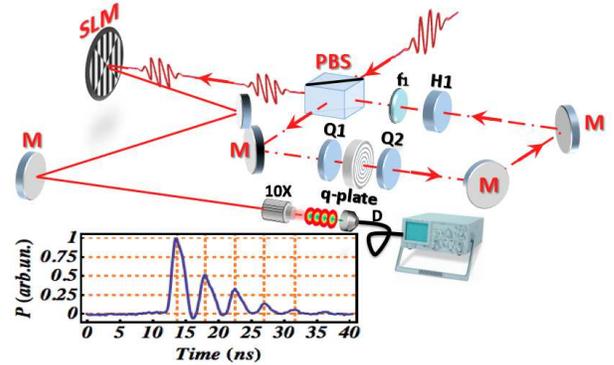}
    \caption{\label{fig:fig1} (Color online) Schematic of our setup. Legend: PBS -- polarizing beam splitter;
H$_1$ -- half-wave plate (HWP); Q$_1$,Q$_2$ -- quarter-wave plates (QWP);
M -- mirror; $f_1$ -- lens. In addition to the main ring cavity setup, an SLM
followed by detection in a single-mode fiber have been used to verify the
OAM content of output pulses for the OAM sequence generation experiments
(as in Fig.\ \protect\ref{fig:fig2}a,b). In the OAM-to-time
conversion experiment we moved the SLM at the input of the ring cavity, in order to
generate a given input OAM, while in the output the single-mode fiber detector was used
for selecting the vanishing OAM output (as in Fig.\ \protect\ref{fig:fig2}c). The inset shows a typical
time sequence of pulses obtained at the output.}
\end{center}
\end{figure}

Central to our approach is the $q$-plate, a device allowing control of OAM by the light
polarization (or photon ``spin'') \cite{marrucci06prl,marrucci11jo}.
A single $q$-plate, however, couples the polarization with two values only of OAM.
To address multiple OAM values one needs to cascade several $q$-plates
in sequence \cite{marrucci06apl,nagali09opex}.
Here, we consider an alternative scheme, as shown in Fig.~\ref{fig:fig1}, in which a
single $q$-plate is inserted in a ring optical cavity, so that a light pulse may
undergo multiple passages through the same $q$-plate, thus addressing
multiple OAM values at different times.
In our experiments, a titanium-sapphire laser operating at 1 kHz repetition rate generated
pulses having a duration of about 130 fs. The ring cavity round trip time was about 4 ns.
A fast photodiode connected to a 500 MHz oscilloscope was used for time-domain
detection (see inset of Fig.~\ref{fig:fig1}). To analyze the setup workings,
let us denote with $\ket{P,\ell}$ an arbitrary
polarization-OAM spinorbit state, where $P$ indicates the polarization,
e.g. $L,R,H,V$ for left-circular, right-circular, horizontal, and vertical polarizations, respectively,
and $\ell$ is an integer specifying the OAM eigenvalue. For simplicity we ignore in the following the radial
mode profile. However, a lens imaging the $q$-plate on itself after a round trip is
needed to erase the radial profile transformations arising from propagation after the OAM
is changed \cite{nagali09opex}. For so-called ``optimal tuning''
\cite{karimi09apl,piccirillo10apl}, a $q$-plate induces the following transformations
in an optical mode having circular input polarization:
\begin{figure*}[t]
\begin{center}
    \includegraphics[width=16.5cm,draft=false]{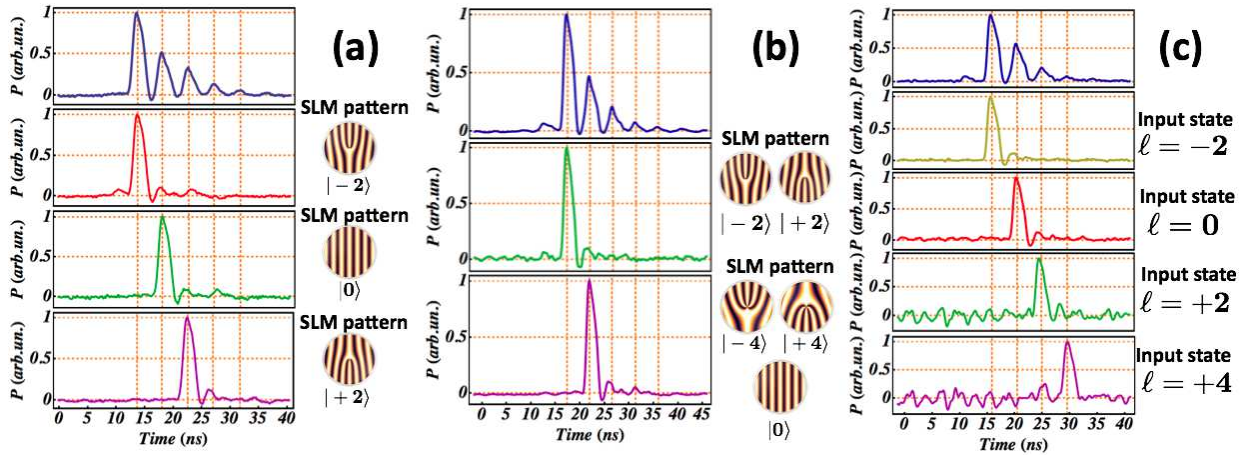}
    \caption{\label{fig:fig2} (Color online) Time traces detected in our experiments. Highest-row panels (a-c):
pulse sequence detected at the exit of the ring cavity when no OAM filtering is applied.
Lower panels, column (a): generation of a sequence of pulses carrying increasing OAM
eigenstates $\ell=-2,0,2,\dots$. The individual pulses detected at different times after
diffraction on the SLM with the displayed
hologram patterns and single-mode optical fiber detection, used for OAM selection. Lower panels,
column (b): generation of a sequence of pulses with superposition states of the following OAM
values: ($-2,2$), ($-4,0,4$), $\dots$. The detection is as in column (a), and in this case different
shown hologram patterns measuring the various OAM eigenstates included in the same superposition return the same pulse timing. Lower panels, column (c): OAM-to-time conversion of the input pulse.
The individual pulses detected after
selecting only zero OAM at the output are shown for different values of the input OAM.}
\end{center}
\end{figure*}
\begin{eqnarray}\label{eq:qplate}
    \ket{L,\ell}&\rightarrow&\ket{R,\ell+2q}\label{eq:qplate1} \nonumber\\
    \ket{R,\ell}&\rightarrow&\ket{L,\ell-2q}\label{eq:qplate2}
\end{eqnarray}
where $q$ is the (integer or half-integer) $q$-plate
topological charge. In our experiments we had $q=1$, but we will
keep the theoretical discussion general. Equations
(\ref{eq:qplate}) show that each passage through the $q$-plate can
be used to shift the OAM value by $\pm 2q$, where the sign depends
on the input polarization handedness (note that a spiral phase-plate could similarly
shift the OAM at each passage, but in a fixed direction). If linear or elliptical
polarizations are used at the $q$-plate input, each passage will give rise to a
superposition state of both the $+2q$ and $-2q$ shifted OAM values.

Let us first discuss the main loop occurring inside the ring cavity. Consider a pulse
with OAM $\ell$ travelling inside the cavity, which has been just reflected by the polarizing
beam-splitter (PBS) used as input/output port. The
PBS is assumed to transmit the $H$ and reflect the $V$ polarization, so we take
the pulse to be in state $\ket{V,\ell}$. The OAM is sign-inverted at each reflection,
but we will ignore this effect that is finally compensated if the total number of reflections
is even. The first quarter-wave plate (QWP) changes the input pulse into the superposition
$\alpha\ket{L,\ell}+\beta\ket{R,\ell}$ where $\alpha=(e^{-i2\theta}+i)/2$
and $\beta=-(e^{i2\theta}+i)/2$, $\theta$ being the angle at which the QWP
optical axis is rotated with respect to the $H$ axis. After the $q$-plate,
the pulse is then cast in the spinorbit entangled state
$\alpha\ket{R,\ell+2q}+\beta\ket{L,\ell-2q}$. The second QWP is oriented
at an angle of $-\pi/4$, so that the pulse state at the end of the round trip,
just before the PBS, is the following: $-i\alpha\ket{V,\ell+2q}+\beta\ket{H,\ell-2q}$.
At the PBS, the $V$ component is reflected and starts a new cycle, while the $H$
component is coupled out of the cavity. The fraction of energy coupled out
at each round trip is given by $\epsilon=|\beta|^2=\left[1+\sin(2\theta)\right]/2$,
which can be adjusted by rotating the first QWP.  In our experiments we set
$\epsilon=0.5$, by taking $\theta=0$. The pulse
reflected at the PBS now has a shifted OAM $\ell\rightarrow\ell+2q$
and starts a new cycle which can be analyzed as the previous one.
The output component has OAM $\ell-2q$. Let us now consider the first cycle
after the input pulse has entered the cavity through the PBS, as this
requires a distinct treatment. In this case, indeed, the
pulse polarization is $H$. Therefore, the coefficients
$\alpha$ and $\beta$ after the first QWP
will be swapped (with a sign change for one), so that at the end of the first round
trip the fraction of energy coupled out is $|\alpha|^2=1-\epsilon$. The second
cycle starts with $V$ polarization and can be analyzed as discussed above.

We consider now some specific applications of our setup. The first is for
generating a sequence of pulses with increasing OAM,
so that OAM and time bin of the pulse are correlated. This is obtained by
entering the ring cavity with a first pulse with a given OAM, for example zero (i.e.,
a normal TEM$_{00}$ gaussian beam). We will then obtain at the output
the increasing-OAM sequence $\ell=-2q,0,2q,4q,6q,\dots$. The first output pulse (with $\ell=-2q$)
will have a fraction $p_1=1-\epsilon$ of the input pulse energy, while the subsequent ones will have an
energy fraction $p_n=\epsilon^2(1-\epsilon)^{(n-2)}$, where $n=2,3,\dots$ is the pulse
number (or time bin) in the sequence. An example of this behavior is shown in
Fig.\ \ref{fig:fig2}a.

The OAM sequence can be also turned into decreasing, instead of increasing, by setting
the second QWP at angle $+\pi/4$. In this case the output OAM sequence (for a gaussian
input) will be $\ell=+2q, 0, -2q, -4q, -6q, ...$. Another possibility is to set the
second QWP to an intermediate angle, which gives rise to the following sequence of OAM
superposition states (the polarization is omitted): $\ket{\psi}_1=\alpha_{-2q}\ket{-2q}+\alpha_{2q}\ket{+2q}$,
$\ket{\psi}_2=\alpha_{-4q}\ket{-4q}+\alpha_0\ket{0}+\alpha_{4q}\ket{+4q}$,
$\ket{\psi}_3=\alpha_{-6q}\ket{-6q}+\alpha_{-2q}\ket{-2q}+\alpha_{2q}\ket{+2q}+\alpha_{6q}\ket{+6q}$,
$\dots$. The superposition coefficients $\alpha_i$ can be partially controlled using the second QWP and
the half-wave plate (HWP). For example, the OAM distribution can be made symmetric ($|\alpha_{-i}|=|\alpha_{i}|$) by setting the second QWP and the HWP to an angle of zero.
An example of this workings regime of our setup
is shown in Fig.\ \ref{fig:fig2}b.

Another possible application of our apparatus is for converting the input information encoded in OAM
into time-bin encoding. The setup is essentially the same as before, but in the output of the
PBS we add an OAM filter to detect selectively $\ell=0$, for example
by coupling the outgoing light into a single-mode fiber (without an intermediate SLM in this case).
Then we operate the loop so as to decrease the OAM at each round trip. If the number of
round-trips needed to observe a pulse in the output is $n$, then the input OAM
is given by $\ell= 2(n-2)q$. In other words, the time delay of the detected pulse is proportional
to the input OAM value. This method is illustrated in Fig.\ \ref{fig:fig2}c. We note that
this method requires prior knowledge of the sign of the input OAM, so as to operate the shifter
in the correct direction. An alternative possibility, when the OAM sign is unknown, is to operate
the setup with the second QWP rotated at angle zero. In this way, the loop will create superposition
states which include both increasing and decreasing $\ell$ values at each iteration,
and the time bin $n$ of the first pulse giving a nonvanishing signal after the OAM filter
will return the absolute value of the input OAM.

The detection efficiency of our setup for a given $\ell$ input is given by the same energy fraction $p_n$
given above for the generation case (not including the OAM filter efficiency). In our experiment we addressed only few OAM values because we set $\epsilon=0.5$, giving rise to a rapid exponential decrease of the
output energy, as $p_n = 0.5^{n}$. Higher OAM values, e.g. corresponding to pulse number $n$,
can be however reached by adjusting the output coupling factor to the optimal value $\epsilon=2/n$ (for $n\ge2$). This returns the optimized efficiency $p_n=4(n-2)^{n-2}/n^n$, which for high $n$
decreases only as $n^{-2}$. When the input pulse is in a superposition of different OAM components,
more pulses are obtained at the output, with amplitudes corresponding to the input OAM power spectrum
except for a rescaling by the detection efficiency $p_n$. Therefore, the input OAM spiral spectrum can be
visualized by a common oscilloscope. The same approach may be applied to single photons, although
in its present form it would be a low efficiency method. The efficiency of our setup could however be
substantially increased by adopting a fast Pockel cell to couple the light in and out, and, for the
OAM-to-time conversion case, by replacing the PBS and subsequent OAM filter with a suitable
hollow mirror, so as to let only the vanishing OAM state out of the cavity.

In conclusion, we reported a novel method for addressing multiple OAM values by coupling with
the time degree of freedom. This approach may find applications in OAM-based quantum
information schemes \cite{nagali09natph,nagali10prl} or in digital spiral imaging and microscopy \cite{furhapter05,torner05}. We acknowledge the
financial support of the FET-Open Program within the 7$^{th}$ Framework Programme of the
European Commission under Grant No. 255914, Phorbitech.

\bibliographystyle{ol}

\end{document}